\documentclass[namedreferences,natbib]{solarphysics}
\usepackage[optionalrh,solaromanenum]{spr-sola-addons} 

\makeatletter
\def\@thedoi{10.1007/s11207-013-0270-3}
\makeatother

\usepackage{graphicx}                    
\usepackage{color}                       
\usepackage{url}                         

\newcounter{IonCS}
\newcommand{\ion}[2]{\setcounter{IonCS}{#2}#1\,{\scshape{\roman{IonCS}}}}
\newcommand{\arcsec}{\mbox{$^{\prime\prime}$}}
\newcommand{\sect}[1]{Section\,\ref{#1}}
\newcommand{\fig}[1]{Figure\,\ref{#1}}
\newcommand{\figs}[1]{Figures\,\ref{#1}}

\newcommand{\eqn}[1]{Equation\,(\ref{#1})}
\newcommand{\bm}[1]{\mbox{\boldmath$#1$\unboldmath}}
\graphicspath{{./}{figs-ps/}{figs/}}

\begin{document}

\begin{article}

\begin{opening}

\title{Magnetic field diagnostics and spatio-temporal variability
       of the solar transition region}

%
\author{H.~\surname{Peter}$^{1}$
       }

\date{Received: 18 December 2012 / Accepted: 4 March 2013}

%
\runningauthor{H. Peter}
\runningtitle{Magnetic field diagnostics and spatio-temporal variability
       of the solar transition region}

%
  \institute{$^{1}$ Max Planck Institute for Solar System Research, 
                    37191 Katlenburg-Lindau, Germany,
                    email: \url{peter@mps.mpg.de}\\
                    $^\star$ The final publication is available at link.springer.com.
                       }

\begin{abstract}
Magnetic field diagnostics of the transition region from the chromosphere to the corona faces us with the problem that one has to apply extreme UV spectro-polarimetry. While for coronal diagnostic techniques already exist through infrared coronagraphy above the limb and radio observations on the disk, for the transition region one has to investigate extreme UV observations. However, so far the success of such observations has been limited, but there are various projects to get spectro-polarimetric data in the extreme UV in the near future. Therefore it is timely to study the polarimetric signals we can expect for such observations through realistic forward modeling.

We employ a 3D MHD forward model of the solar corona and synthesize the Stokes $I$ and Stokes $V$ profiles of \ion{C}{4} (1548\,{\AA}). A signal well above 0.001 in Stokes $V$ can be expected, even when integrating for several minutes in order to reach the required signal-to-noise ratio, despite the fact that the intensity in the model is rapidly changing (just as in observations). Often this variability of the intensity is used as an argument against transition region magnetic diagnostics which requires exposure times of minutes. However, the magnetic field is evolving much slower than the intensity, and thus when integrating in time the degree of (circular) polarization remains rather constant. Our study shows the feasibility to measure the transition region magnetic field, if a polarimetric accuracy on the order of 0.001 can be reached, which we can expect from planned instrumentation.

\end{abstract}

%

\end{opening}

\section{Introduction\label{S:intro}}

There is a general consensus that the outer atmosphere of the Sun (and other cool stars) is heated by one or several mechanisms related to the magnetic field \citep[\textit{e.g.}][]{Schrijver+Zwaan:2000}. Despite the pivotal importance of the magnetic field for our understanding of the corona, actual measurements of the magnetic field are scarce --- mostly we have to rely on extrapolations of the magnetic field from the photosphere \citep[\textit{e.g.},][]{deRosa+al:2009.full}. Because the extrapolations are based on assumptions that might or might not be fulfilled in regions of interest in the corona, there is the need to actually measure the coronal magnetic field. 

Some measurements in the corona have been performed in active region loops above the limb using the Zeeman effect for infrared coronagraphic observations \citep{Lin+al:2000}. These, as well as radio measurements \citep[\textit{e.g.}][]{White:2005}, suffer from low spatial and temporal resolution. A very promising project for diagnostics of the magnetic field in the upper chromosphere is the rocket experiment chromospheric Lyman-alpha spectro-polarimeter \citep[CLASP;][]{Kobayashi+al:2012.full}, planned to be flown in Dec\,2014. This is based on diagnostics using the Hanle effect wiin Ly-$\alpha$ \citep{Trujillo-Bueno+al:2011}.

The first attempt to measure the magnetic field on the solar disk at high resolution in the transition region from the chromosphere to the corona was done using the \textit{Solar Maximum Mission} (SMM) ultraviolet spectrometer and polarimeter \citep[UVSP;][]{Woodgate+al:1980.full}. Reaching a polarimetric accuracy of just below 1\% in the \ion{C}{4} line at 1548\,\AA\ \citep{Henze+al:1982.full,Hagyard+al:1983.full} gave no conclusive results, except maybe in sunspots \citep{Lites:2001}. As it will become clear in this manuscript, an accuracy of 0.1\% would be needed to get useful information on the magnetic field with the \ion{C}{4} line. This accuracy is provided by the solar ultraviolet magnetograph investigation \citep[SUMI;][]{West+al:2004.full} that has been flown on a rocket twice. The \ion{C}{4} data from the latest flight in summer 2012 await calibration and analysis.

The present study will present a forward model which provides synthesized polarimetric data as they are recorded, \textit{e.g.},\ by SUMI. This will allow a direct comparison between model and observations and will (hopefully) provide some guidance for the interpretation of the acquired polarimetric observations. A 3D model is prerequisite to investigate the transition region, because of its highly complex spatial structuring \citep[\textit{e.g.}][]{Peter:2000:sec,Peter:2001:sec}. Consequently, this study is based on a 3D magnetohydrodynamic (MHD) numerical experiment that provides temperature, density, velocity and of course the magnetic field in the corona above a small active region. This model produces a loop-dominated corona \citep{Gudiksen+Nordlund:2002,Gudiksen+Nordlund:2005a,Gudiksen+Nordlund:2005b}. In a statistical sense it reproduces various observational properties \citep{Peter+al:2004,Peter+al:2006}, in particular the persistent transition region redshifts \citep{Peter+Judge:1999,Peter:1999full}. Based on the success of these models the present study goes one step further to investigate not only the profiles of the emission lines, but also the circular polarization due to the Zeeman effect. Another forward modeling investigation of the polarization of the emergent radiation in an MHD model of the extended solar atmosphere, but for the (optically thick) hydrogen Ly-$\alpha$ line, has been carried out by \cite{Stepan+al:2012} paying particular attention to the linear polarization produced by scattering processes and the Hanle effect.

Once the instruments provide the required polarimetric sensitivity, we will have to perform a proper interpretation of the data. The extreme UV lines formed in the transition region (or in hotter parts of the atmosphere) originate from a spatially complex volume, comparable to highly corrugated surfaces. This is a fundamental difference to photospheric magnetic field observations, where the height of the source surface remains at a roughly constant altitude, within one barometric scale height. This more complex source region of transition region lines complicates possible inversions enormously. Furthermore, the high temporal variability of the intensity will be a major problem for the interpretation of the polarimetric data. The forward-model approach presented in this study provides insight into these problems by accounting for all the spatial and temporal complexity.

In \sect{S:synthetic.spectra} the basic concept of the 3D MHD model and the spectral synthesis including the calculation of the Stokes $V$ profile will be introduced. Based on this in \sect{S:diagnostics} we will present how to synthesize the observable quantities and present some sample Stokes $V$ profiles in \sect{S:profiles}. The observational requirements for the exposure times will be discussed in \sect{S:variability} along with the spatio-temporal variability found in observations and in synthesized model data. In \sect{S:real.obs} we will construct a realistic Stokes $V$ observation of \ion{C}{4} and show that we can afford comparably long exposure times for the observations --- and why. Finally, in \sect{S:inversion} we will discuss some simple inversions and their reliability, before we conclude the paper with 
\sect{S:conclusions}.

\section{Synthetic spectra from 3D MHD model\label{S:synthetic.spectra}}

In general, the state of polarization of the light can be described by the Stokes vector $(I,Q,U,V)$. Stokes $I$ represents the integral over all polarization states, while Stokes $V$ is the difference of right- and left-circularly polarized light, and hence carries information on the longitudinal component of the magnetic field through the Zeeman effect. For an observation near disk center and assuming that the magnetic field will be predominantly vertical, we can expect Stokes $Q$ and $U$, which characterize the linear polarization, to be much weaker than Stokes $V$, just as found in photospheric observations in the visible. Because it will turn out that already the Stokes $V$ signal will be at the edge of observability, this study will concentrate on Stokes $I$ and $V$ only.

\subsection{3D MHD model\label{S:model}}

In order to calculate the Stokes profiles, one needs the temperature $T$, density $n$, velocity vector $\bm{v}$, and magnetic field vector $\bm{B}$. These are provided by a 3D MHD model that solves for the induction equation, the conservation of mass, and the momentum and energy balance. Most importantly, the energy balance has to include heat conduction and radiative losses. This is pivotal to set the proper coronal pressure and therefore a prerequisite to synthesize coronal emission lines which are very sensitive to the temperature and density.

The MHD model used for this study has been published already by \cite{Gudiksen+Nordlund:2002,Gudiksen+Nordlund:2005a,Gudiksen+Nordlund:2005b}, and an analysis of this model in terms of Doppler shifts and emission measure and a comparison to observations was presented by \cite{Peter+al:2004,Peter+al:2006}. In the model the plasma is heated through Ohmic dissipation of currents that are induced by the braiding of magnetic field lines through the horizontal photospheric granular motions. The good match to the observations showed that this model provides a realistic way to describe the corona in an active region, accounting for the spatial and temporal variability. Since then further models of similar type solidified these results discussing details of the heat input \citep{Bingert+Peter:2011,Bingert+Peter:2013}, providing further insight into the persistent transition region Doppler shifts \citep{Hansteen+al:2010,Zacharias+al:2011.doppler}, transient events in the corona \citep{Zacharias+al:2011.blob}, or the constant cross section of loops \citep{Peter+Bingert:2012}. All these results give good confidence that this type of model will also allow for a reliable and realistic determination of the Stokes $V$ profiles in a transition region extreme UV line.

\subsection{Intensity spectra: Stokes $I$ profiles\label{S:intensity}}

To calculate the Stokes profile $I(\lambda)$ we follow exactly the procedure of \cite{Peter+al:2004,Peter+al:2006}. Under the assumption of ionization equilibrium we calculate the emissivity (energy loss per time and volume) at each grid point in the computational domain in the \ion{C}{4} (1548\,\AA) line using the {\sc{Chianti}} atomic data base \citep{Dere+al:1997,Young+al:2003}. In order to avoid aliasing effects we interpolate the mesh in the vertical direction. The line width is assumed to be the thermal width, $\Delta\lambda_{\rm{D}}$, and the Doppler shift is given by the line-of-sight component of the velocity vector (here the vertical component). This provides a line profile $I(\lambda)$ at each grid point.

\subsection{Stokes $V$ profiles: weak-field approximation\label{S:stokes}}

In a magnetic field a spectral line will be affected by the Zeeman effect. The Zeeman splitting is given through
\begin{equation}
\Delta\lambda_B = \frac{e}{4\,\pi\,c\,m_{\rm{e}}}~
            \bar{g}\,\lambda_0^2\,B_{\rm{los}},
\end{equation}
with the elementary charge $e$, the speed of light $c$, the electron mass $m_{\rm{e}}$, the rest wavelength $\lambda_0$, the effective Land\'e factor $\bar{g}$ and the component of the magnetic field along the lone of sight $B_{\rm{los}}$.

In the weak-field limit the Stokes $V$ signal is given through through a Fourier expansion up to first order of the Stokes $I$ line profile \citep[\textit{e.g.},][Section\,11.9]{Stenflo:1994},
\begin{eqnarray}
\label{E:stokes}
V(\lambda)  \approx  -\Delta\lambda_B 
            \frac{\partial I}{\partial\lambda}
   ~\approx~  -4.67{\cdot}10^{-13}~
              \bar{g}\,(\lambda_0[{\rm{\AA}}])^2\,B_{\rm{los}}[{\rm{G}}]~
              \frac{\partial I}{\partial\lambda[{\rm{\AA}}]}     .
\end{eqnarray}

In this study we will concentrate on the \ion{C}{4} line at 1548\,\AA\ when investigating the transition region magnetic field, as it has a high diagnostic potential: it is a strong line at a long wavelength $\lambda_0$ (for transition region extreme UV lines) with a decent effective Land\'e factor $\bar{g}$. This combination provides the best potential among the transition region lines.

The effective Land\'e factor for the 1548\,\AA\ line 
($^{2}{\rm{P}}_{3/2} \,{\to}\, ^{2}{\rm{S}}_{1/2}$) is $\bar{g}{=}7/6{\approx}1.167$ \citep[for details on the calculation of $\bar{g}$ see \textit{e.g.}][Sects.\,6.4 \& 6.5]{Stenflo:1994}. While the 1550\,\AA\ line of the \ion{C}{4} doublet has a slightly higher effective Land\'e factor ($\bar{g}{=}4/3$), its radiance is lower by a factor of about two, which is why we concentrate on the 1548\,\AA\ line.

Even for a very strong magnetic field in the transition region of 1000\,G the splitting for the \ion{C}{4} lines would be below $\Delta\lambda_B{<}0.15$\,pm corresponding to less than 0.3\,km\,s$^{-1}$ in Doppler shift units. This can be considered as an upper limit.
Consequently, in the case of \ion{C}{4} the Zeeman splitting is much smaller than the Doppler broadening of the line, $\Delta\lambda_B{\ll}\Delta\lambda_{\rm{D}}$, the latter being about 6\,pm corresponding to 12\,km\,s$^{-1}$ at the line formation temperature of \ion{C}{4} of about $10^5$\,K. Thus the application of the weak-field limit for \eqn{E:stokes} is justified.

From \eqn{E:stokes} we can roughly estimate the Stokes $V/I$ signal to be expected in \ion{C}{4}. For this we assume a magnetic field of 100\,G (above a pore or a strong network patch) in the transition region some 3\,Mm above the photosphere.

Then, for a line width of the order of the thermal width the expected signal is about $V/I\approx0.002$, which would be measurable with planned instrumentation (cf.\ \sect{S:conclusions}).

With the line-of-sight magnetic field from the 3D model and the Stokes $I$ profile from \sect{S:intensity} we can compute the Stokes $V$ profile according to \eqn{E:stokes} at each grid point on the interpolated mesh of the computational domain.

In this study we will concentrate on the \ion{C}{4} line formed
at about $10^5$\,K. It could be easily also done for any other extreme UV
emission line, \textit{e.g.}, the \ion{O}{6} line at 1032\,\AA\ formed at some 300\,000\,K,
or the \ion{Mg}{10} line at 625\,\AA\ formed at $10^6$\,K to study the coronal
magnetic field. However, at the shorter wavelengths the Stokes $V$ signal
will be weaker. Furthermore, at higher temperature, on average found at higher
heights, the magnetic field will be weaker. Thus to detect a Stokes $V$ signal
from these lines the sensitivity of future instruments would have to be well
below 0.1\%.

\begin{figure}
\centerline{\includegraphics{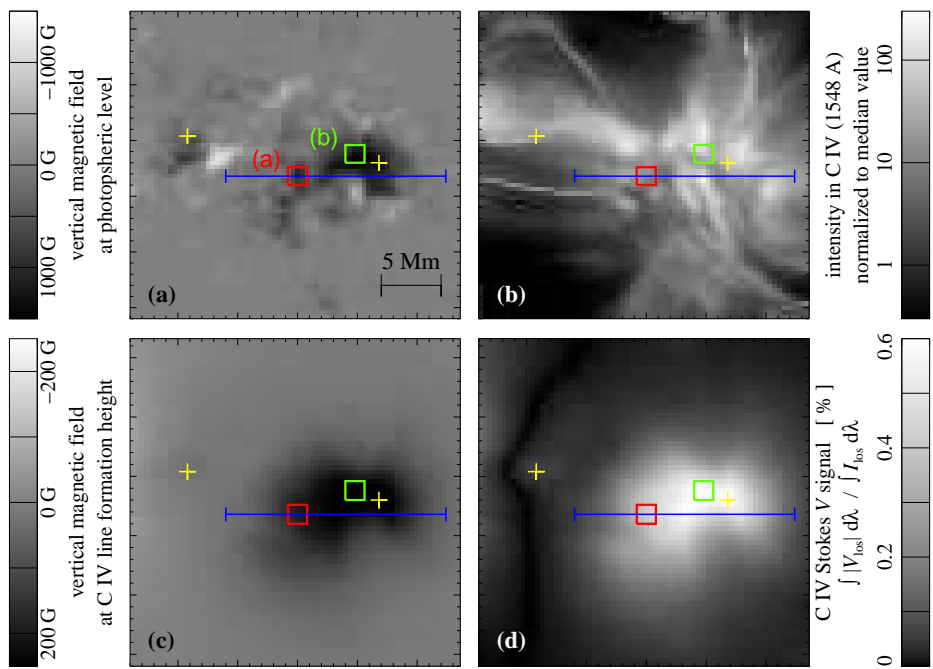}}
\caption{Snapshot of a view when looking from straight above at the computational box (along the vertical direction). Panel (a) shows the line-of-sight magnetic field at the bottom of the box defined through the boundary condition. Panel (b) displays the synthesized emission of the \ion{C}{4} line, integrated along the line of sight. The line-of-sight component of the transition region magnetic field at the height of the formation of \ion{C}{4} is seen in panel (c). Panel (d) shows the resulting integrated Stokes $V$ signal normalized by the line intensity (see Eq.\,\ref{E:stokes.map}).
The field of view shown here is about 27\,${\times}$\,27\,Mm$^2$ and represents part of the whole computational domain.
The two yellow crosses denote the locations for which the sample Stokes $V$ spectra are plotted in \fig{F:stokes_sample}a.
The red and green boxes show the regions over which realistic Stokes profiles are synthesized (\fig{F:stokes.profiles}) and in which the temporal variation is analyzed (\fig{F:stokes.vari}). The long blue bar indicates the line along which the spatial variation is analyzed (\fig{F:stokes.spatial}).}
\label{F:stokes.maps}
\end{figure}

\section{Synthetic observation and diagnostics}\label{S:diagnostics}

For this study we will restrict the discussion to a part of the computational domain covering about 27\,${\times}$\,27\,Mm$^2$ in the horizontal directions. This region of interest contains a magnetic structure at the bottom boundary that resembles a pore on the real Sun. The (vertical component of) the magnetic field at the bottom boundary, \textit{i.e.} the photosphere, in this area is shown in \fig{F:stokes.maps}\,(a). In this region the photospheric magnetic field is predominantly of one sign.

We will investigate synthetic extreme UV observations for which the line of sight is aligned with the vertical. For this we integrate the Stokes $I(\lambda)$ and $V(\lambda)$ profiles along the vertical direction,
\begin{equation}
I_{\rm{los}}(\lambda) = \int I(\lambda)~{\rm{d}}z
\qquad;\qquad
V_{\rm{los}}(\lambda) = \int V(\lambda)~{\rm{d}}z ~.
\end{equation}
This will produce observables as they would be found in actual observations of the Sun close to disk center when employing an extreme UV spectro-polarimeter. These data can then be analyzed in the same way as actual observations of the Sun would be handled. For instance, one can derive maps of the total line intensity, Doppler shifts, or the wavelength-integrated Stokes $V$ signal.
In particular, we define the total intensity and the integrated Stokes $V$ signal as
\begin{equation}\label{E:stokes.map}
\overline{I\,} = {\int\, I_{\rm{los}}(\lambda) ~ {\rm{d}}\lambda}
\qquad;\qquad
overline{V} = \frac{1}{~\overline{I\,}~}
               \int\, |V_{\rm{los}}|(\lambda) ~ {\rm{d}}\lambda~ 
               ~.
\end{equation}
The latter gives the fraction of the (unsigned) Stokes signal along the line of sight when compared to the total line emission. In \figs{F:stokes.maps}b and \ref{F:stokes.maps}d we show the resulting maps of $\overline{I\,}$ and $\overline{V}$ for the part of the computational domain under investigation here.

\begin{figure}
\parbox{0.6\textwidth}{
\includegraphics[width=0.6\textwidth]{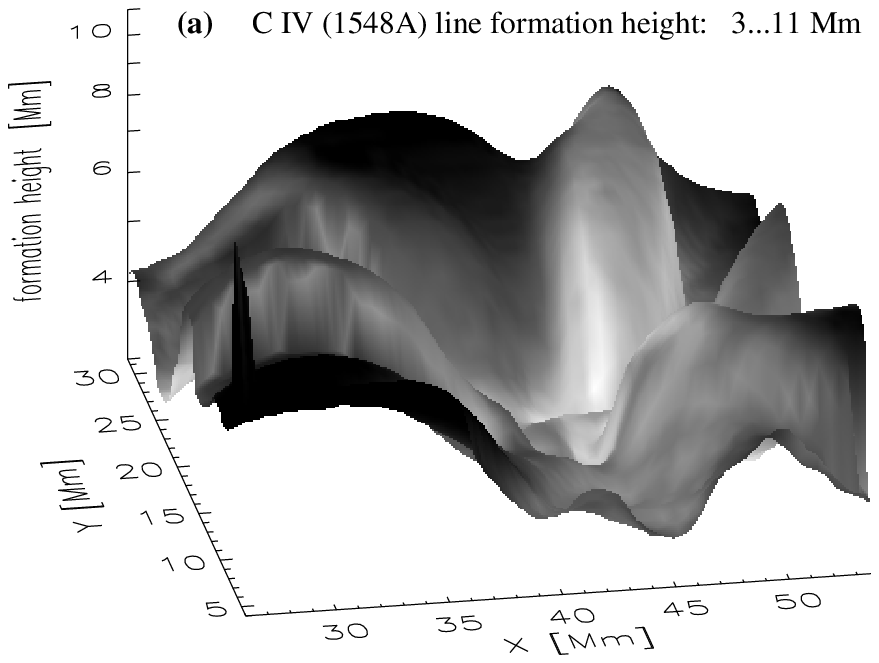}} \hfill 
\parbox{0.36\textwidth}{
\includegraphics[width=0.36\textwidth]{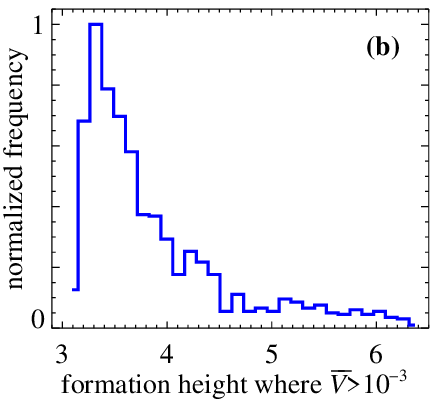}}
\caption{Line-formation height of \ion{C}{4}. The surface in panel (a) shows the height where \ion{C}{4} is originating. Displayed on that surface is the emission from \ion{C}{4} to show the (inverse) relation of emission and formation height. The horizontal extent of the field of view is the same as in \fig{F:stokes.maps}. Panel (b) presents the distribution of formation heights in those areas where the Stokes signal $\overline{V}$ as defined in \eqn{E:stokes.map} exceeds $10^{-3}$. }
\label{F:stokes.height}
\end{figure}

Having access to the three-dimensional distribution of the \ion{C}{4} emission in the computational domain, at each horizontal location $(x,y)$ we can derive at which height $z$ the main contribution of \ion{C}{4} is originating from. This defines the line-formation height, which is visualized in \fig{F:stokes.height}a. As noted before, this is a highly corrugated surface. Typically, the line-formation height is the lowest where the intensity is the highest -- where the heating is particularly high the transition will move to lower heights and produce more emission.
Even when considering only regions with considerable field strength, where the Stokes signal $\overline{V}$ as defined in \eqn{E:stokes.map} exceeds $10^{-3}$, there is a wide range of heights of the source region of \ion{C}{4}. The histogram in \fig{F:stokes.height}b shows that there the formation of \ion{C}{4} is mostly found between 3\,Mm and 4\,Mm, sometimes reaching up to 6\,Mm. This corresponds to several chromospheric pressure scales height ($\approx$300\,km).

Having the line-formation-height, we can now extract the magnetic field at the source region of \ion{C}{4}. We plot the vertical component of this, \textit{i.e.}, the line-of-sight component, in \fig{F:stokes.maps}c. This map of the transition region magnetic field is \emph{not} just a horizontal cut, but shows the magnetic field at the actual height of the main transition region emission for each horizontal location, \textit{i.e.}, on the corrugated surface shown in \fig{F:stokes.height}a.

This surface of the maximum contribution of \ion{C}{4} coincides with the location where the temperature jump into the corona is found. However, the spatial structure of the transition region is even more complicated than this. The emission of the optically thin \ion{C}{4} line is not restricted to this surface, but pockets of $10^5$\,K cool plasma are also found higher up in the (generally) hotter volume of the corona, as can be seen from the images and movies of \cite{Peter+al:2006}. One should keep this in mind when interpreting the data -- along the line-of-sight various structure can contribute to the emission seen in transition region lines.

 Comparing the photospheric and transition region magnetic field in \figs{F:stokes.maps}a and \ref{F:stokes.maps}c, it is clear that the transition region field is much smoother and no longer shows any signs of (small-scale) mixed polarities. The opposite polarities in the photosphere are a couple of Mm apart, so at a height of 3 Mm and higher, where \ion{C}{4} is formed, all these mixed polarities are closed already. Similar to a simple potential field expansion, the 3D MHD model shows the expansion and smoothing of the magnetic field with height.

\section{Stokes $V$ line profiles\label{S:profiles}}

\begin{figure}
\centerline{\includegraphics{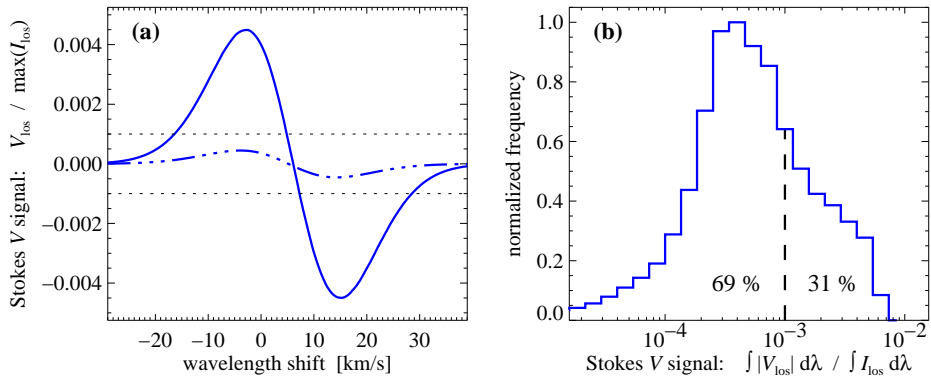}}
\caption{%
Synthesized Stokes $V$ profiles and histogram of the distribution of Stokes $V$ signals.
Panel (a) shows two sample Stokes $V$ profiles normalized to the peak line intensity for the locations marked by crosses in \fig{F:stokes.maps} (the stronger signal shown as a solid line corresponds to the right cross). 
Panel (b) displays the distribution of Stokes $V$ signals as defined in \eqn{E:stokes.map} for the field of view shown in \fig{F:stokes.maps}.
About one third of the area shows a Stokes $V$ signal above 0.1\,\%. Alternatively one could also plot the histogram of the peak values of the Stokes $V$ profiles, which would show a very similar result.
See \sect{S:profiles}.}
\label{F:stokes_sample}
\end{figure}

Two examples of Stokes $V$ profiles are shown in \fig{F:stokes_sample}a. The sample from the location close to the center of the magnetic field concentration in the photosphere shows a peak-to-peak signal in Stokes $V$ of \ion{C}{4} normalized to the peak intensity of almost 1\%. This profile is among the strongest Stokes $V$ signals in the field-of-view (right cross in \fig{F:stokes.maps}).
The signal is symmetric and close to a Stokes $V$ signal in an idealized situation. In regions where the magnetic field is weak, but where still significant emission in \ion{C}{4} is present (left cross in \fig{F:stokes.maps}), the Stokes $V$ signal typically is weaker than 0.1\%.

The distribution of the \ion{C}{4} Stokes $V$ signals in the field-of-view displayed in \fig{F:stokes_sample}b reveals that about one third of the area shows a signal above 0.1\%. This is within the detection limit of the SUMI rocket \citep{West+al:2004.full}. Likewise, the future rocket mission CLASP \citep{Kobayashi+al:2012.full} will achieve this accuracy in the extreme UV (albeit at Ly-$\alpha$). Furthermore the space-based observatory SolmeX that was proposed to ESA \citep{Peter+al:2012.solmex.full} would have included a spectro-polarimeter with the required capabilities for \ion{C}{4} observations (see also \sect{S:conclusions}). 

The majority of the area in the field-of-view visible in \fig{F:stokes.maps} shows signals much smaller than 0.1\% (cf. \fig{F:stokes_sample}b) that will not be detectable. Mostly these weak signals originate from regions with very low emission in \ion{C}{4}. So besides the weak Stokes $V$ signal also the low intensity would prevent the detection of a signal here because of the limitation in signal-to-noise.

The distribution in \fig{F:stokes_sample}b shows the Stokes signal as defined in \eqn{E:stokes.map}, \textit{i.e.}, the total unsigned Stokes $V$ normalized by the total intensity. The distribution for the peak values of Stokes $V$ normalized by the peak intensity would show a very similar distribution.

\section{Variability of the transition region intensity and magnetic field\label{S:variability}}

\subsection{Observational requirements for exposure times\label{S:required.exp}}

Based on the discussion of \figs{F:stokes.maps}d and \ref{F:stokes_sample}b in \sect{S:profiles}, and considering the order-of-magnitude arguments in \sect{S:stokes} it is clear that instruments would have to detect a Stokes $V$ signal (normalized to $I$) of 0.001 or better. This implies that the signal-to-noise ratio of the recorded data has to better than 1000. Assuming Poisson statistics this leads to the conclusion that the detector has to acquire at least $10^6$ counts (or $10^7$ counts for a $3\sigma$ detection). This will set a limit of the required exposure time.

As a simple experiment one can investigate a time series of actual observations to get an estimate of the exposure time. The most recent instrument that recorded high-quality solar data in \ion{C}{4} is SUMER \citep{Wilhelm+al:1995.full}. Because of the lack of a well suited time series in \ion{C}{4} here data from \ion{Si}{4} are used, a line that forms at similar temperatures as \ion{C}{4}. The two lines share the major observational properties, \textit{e.g.}, variability, line shifts, contrast, etc.

\begin{figure}
\centerline{\includegraphics[width=0.6\textwidth]{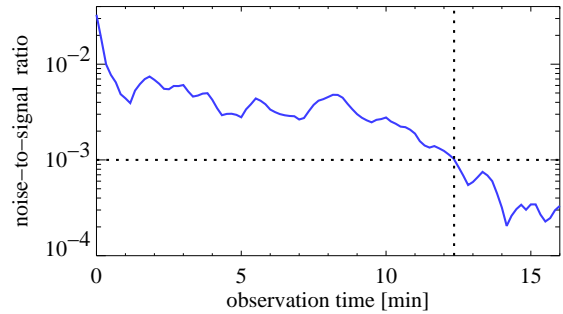}}
\caption{%
Noise-to-signal ratio for a photon-counting detector (SUMER/SOHO) from an actual observation.
The data show \ion{Si}{4} at 140.3\,nm observed in the quiet Sun network. 
After $\approx$12 minutes 10\/$^6$ counts were reached and the noise level dropped to 10$^{-3}$. See \sect{S:required.exp}.}
\label{F:intensity.noise}
\end{figure}

To investigate the noise level, a time series with high temporal cadence is analyzed and the acquired counts are accumulated. This is possible with a photon counting device as SUMER. Assuming Poisson statistics, which is a good approximation for the SUMER detector, the noise-to-signal ratio is derived. This is simply the inverse of the square root of the accumulated counts. In \fig{F:intensity.noise} this noise-to-signal ratio based on the accumulated counts is shown as a function of the observational time. As expected the noise level generally drops in time. The details depend on the time-variable emission, of course. For the particular example shown in \fig{F:intensity.noise} after some 12 minutes a noise level of 0.001 is reached. 

Therefore, to reach a signal-to-noise level to detect a Stokes $V$ signal of 0.001 one would have to integrate for some 12 minutes, too. The data shown here are for a bright patch in the network, so for regions with higher magnetic field, \textit{e.g.}, in the vicinity of a pore, we can expect higher emission in the transition region lines, and thus a shorter exposure time would be sufficient. Also a more modern instrument could have a higher throughput and larger aperture, which would in part be counterweight by the additional optics needed to analyze the state of polarization of the incoming light. So as a very rough estimate one might need exposure times of the order of minutes.

In the light of the high variability of the transition region emission this poses the question if Stokes $V$ observations with such comparably long exposure times are meaningful. After presenting the transition region variability in actual observations (\sect{S:sumer.obs}) and in the synthesized model data (\sect{S:synt.vari}) the discussion will turn to this question in \sect{S:real.obs}.

\subsection{Actually observed transition region variability\label{S:sumer.obs}}

\begin{figure}
\centerline{\includegraphics{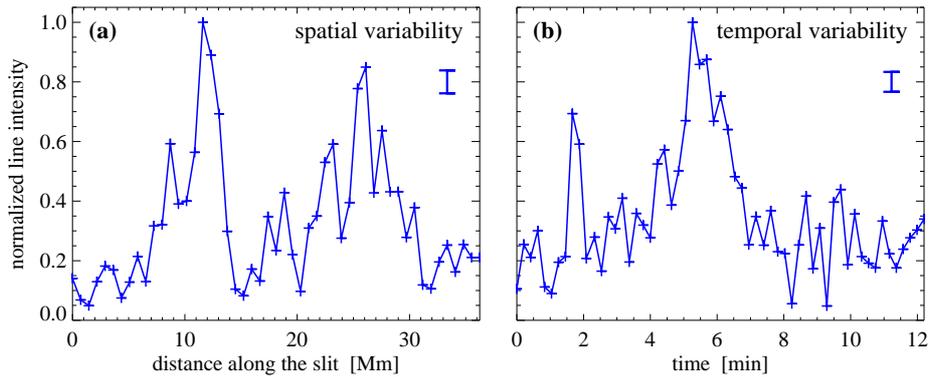}}
\caption{Observed actual spatial and temporal variability of the transition region.
  Example of SUMER data showing the intensity of the \ion{O}{6} line at 1038\,{\AA} 
  observed on 16 May 1998. They represent a region with strong network activity.
  The left panel (a) displays the spatial variation (along the slit),
  with a pixel size of 1{\arcsec} (crosses), \textit{i.e.}\ at the highest
  possible spatial sampling, at an exposure time of 10\,s.
  The right panel (b) shows the temporal variation in a single
  1{\arcsec}${\times}$1{\arcsec} spatial pixel (crosses).
  The bars in the panels indicate the average uncertainty for the
  line intensity (Poisson noise).
  The crosses indicate the spatial pixels (left) and times of exposures (right).
  The temporal and the spatial variability is down to the resolution limit
  of SUMER.
  See \sect{S:sumer.obs}.}
\label{F:obs.variability}
\end{figure}

Just as an example, \fig{F:obs.variability} shows the temporal and spatial variability of the transition region emission. The spectra were acquired in the quiet Sun in a region with strong network activity. The spatial variation (panel a) shows a 35\,Mm ($\approx$48\arcsec) long cut through two network patches with enhanced emission. Besides this spatial variation on a scale of some 10\,Mm, also variations down to the resolution limit (1\arcsec\ spatial sampling) can be seen. It has to be emphasized that this pixel-to-pixel variation is \emph{not} noise, but represents real variability not fully resolved by the instrument.

For the temporal variation (\fig{F:obs.variability}b) one can identify fluctuations also down to the sampling of the instrument (here 10\,s). Again, these fast fluctuations are real, and consistent with the very short cooling times in the transition region. Stronger fluctuations well above a factor of two occur in time scales of minutes that are omnipresent on the Sun. In this example two cases, one shorter and one longer can be identified. These brightenings can be classified as blinkers that have been studied abundantly \citep{Harrison+al:1999,Harrison+al:2003,Peter+Brkovic:2003,Brkovic+Peter:2004}

This strong variability in time (and space) raises the question to what extent spectro-polarimetric data will be useful to investigate the transition region magnetic field with instruments that will have limited spatial and temporal resolution. This can be explored by observations synthesized from a model, where the magnetic field is known.

\subsection{Spatio-temporal variability in synthesized observations\label{S:synt.vari}}

\begin{figure}
\centerline{\includegraphics{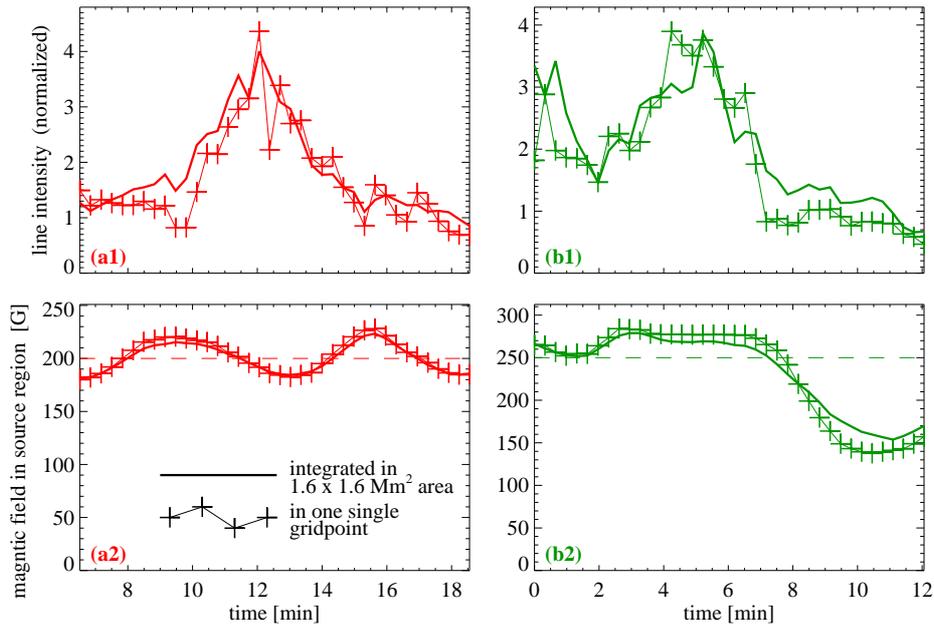}}
\caption{Temporal variability of synthesized transition region emission and magnetic field.
The top panels (a1) and (b1) show the temporal variation in the two 1.6\,${\times}$\,1.6\,Mm$^2$ squares labeled in \fig{F:stokes.maps}a as (a) and (b). The thick solid lines show the variation of the average intensity in the respective square, while the thin lines with the crosses show the variation at the central grid point in that square.
The bottom panels (a2) and (b2) show the variation of the magnetic field at the line formation height of \ion{C}{4}, \textit{i.e.}, the source region of the transition region emission. Again, this is done for the average within the respective square (thick) and the central grid point (thin with crosses). The dashed lines indicate the magnetic field inverted from the average Stokes $V$ signal as discussed in \sect{S:inversion}.
See \sect{S:synt.vari}.}
\label{F:stokes.vari}
\end{figure}

Examples for the temporal variability of the transition region emission synthesized from the 3D MHD model are shown in \fig{F:stokes.vari} (top panels) for the squares labeled (a) and (b) in \fig{F:stokes.maps}. These regions are selected to be in regions near the main polarity with medium (a) and high (b) intensity in \ion{C}{4}. In the figure the variation integrated over an area of 1.6$\times$1.6\,Mm$^2$ (roughly 2\arcsec$\times$2\arcsec) is plotted along with the intensity of the simulation in the center of that square. As expected, the variation integrated over the square is smoother than the one in its center, with the variation being quite similar in both cases.

In both examples shown in \fig{F:stokes.vari} transient brightenings are visible that last a few minutes with an enhancement of the intensity by about a factor of about four. These brightenings are induced by short increases of the heating rate, which are ubiquitously present in these 3D MHD models \citep[see also][]{Bingert+Peter:2011,Bingert+Peter:2013}. The brightenings shown here have properties similar to the observed ones in \fig{F:obs.variability}, and could therefore be considered as a valuable model for blinkers \citep{Harrison+al:1999,Harrison+al:2003}. However, the relation of the observed blinkers to the brightenings in these 3D MHD models will be left for a future study.

In contrast to the spectroscopic observations, in the model one can compare the intensity variation to the magnetic field $B_{\rm{TR}}$ in the source region of the transition region emission (bottom row of \fig{F:stokes.vari}). It is clear that the magnetic field strength in the source region changes only slightly, while the intensity changes dramatically.
The fact that the plasma-$\beta$ parameter is much smaller than unity in the source region of \ion{C}{4} \citep[cf.][their Figure\,12]{Peter+al:2006} allows very strong and rapid intensity variation while the magnetic field remains mostly unaffected. This opens the possibility to infer the transition region magnetic field despite the high temporal variability of the emission (cf.\ \sect{S:real.obs}).

In example (a) in the left column the magnetic field strength changes only by some 20\%, while the intensity increases by a factor of three. With the increased heating the transition region moves downward to higher densities to be able to radiate the increased amount of heat input. Because the scale height in the chromosphere is only some 300\,km, moving down by about 150\,km is sufficient to relocate the transition region to densities a factor of about 1.7 higher. The emission goes with the density squared and thus this change of height of the source region by 150\,km will change the intensity (and thus the radiative losses) by a factor of three. Over this small change in height the magnetic field is changing only slightly, hence the small decrease of $B$ by some 30\,G at the time of the peak of emission in the example (a). 

The example (b) in the right column of \fig{F:stokes.vari} also shows only a small change in magnetic field while the intensity changes significantly. Here the magnetic field in the area is dropping towards the end of the time frame shown because of the (comparably slow) changes induced by the footpoint motions. This leads to weaker currents, less heating and consequently less transition region emission.

In \fig{F:stokes.vari} only ${\int}I_{\rm{los}}\,{\rm{d}}\lambda$ and the magnetic field in the source region, $B_{\rm{TR}}$ are plotted. Because according to \eqn{E:stokes} Stokes\,$V$ is basically proportional to $I$ and $B_{\rm{TR}}$, and because $B_{\rm{TR}}$ is quite constant, Stokes\,$V$ is changing in a manner very similar to $I$. This is why ${\int}|V_{\rm{los}}|\,{\rm{d}}\lambda$ is not plotted in \fig{F:stokes.vari}. 
Normalizing ${\int}|V_{\rm{los}}|\,{\rm{d}}\lambda$ by ${\int}I_{\rm{los}}\,{\rm{d}}\lambda$, \textit{i.e.} $\overline{V}$ as defined in \eqn{E:stokes.map}, basically gives the magnetic field, see also \sect{S:inversion} and \eqn{E:magnetograph}. Therefore the variation of the (normalized) Stokes $V$ signal, $\overline{V}$, closely follows the magnetic field in the source region, $B_{\rm{TR}}$. Thus we plotted only $B_{\rm{TR}}$ but not $\overline{V}$ in \fig{F:stokes.vari}.

\begin{figure}
\centerline{\includegraphics{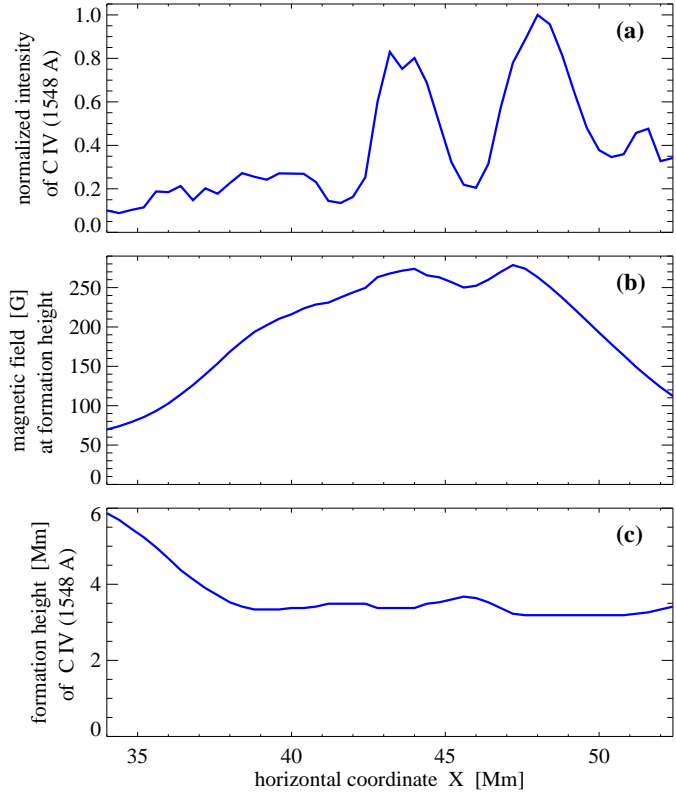}}
\caption{Spatial variation of the synthesized transition region emission, magnetic field and line formation height. The spatial variation is plotted along the long blue horizontal bar marked in in \fig{F:stokes.maps} and is shown for the same time as the snapshot displayed there.
See \sect{S:synt.vari}.
}
\label{F:stokes.spatial}
\end{figure}

The spatial variation basically shows the same properties as the temporal variation: strong intensity variation with only small changes of the magnetic field in the source region. To illustrate this, \fig{F:stokes.spatial} shows the variation along the blue bar in \fig{F:stokes.maps}. Despite the large-scale variation of the magnetic field across the region of the pore, in the region of strongest magnetic field ($x$ from about 40\,Mm to 50\,Mm) there is only a small variation in magnetic field, while the \ion{C}{4} intensity changes by a factor of 4 to 5. Here the close relation between the decreased emission and magnetic field near $x\approx46$\,Mm to the increase in formation height can be seen. The small increase in formation height (due to less heating) by some 300\,km leads to only a small change in magnetic field. Because the (chromospheric) scale height is comparable to this change in formation height, the change in emission is quite dramatic.

\section{A realistic synthetic observation: can we afford long exposure times?\label{S:real.obs}}

The above discussion shows that the observations and the synthesized emission from the 3D model show very strong variations in the emissivity. A future spectro-polarimetric instrument would have only a limited resolution in space \emph{and} time because of the limitations in count rate and signal-to-noise to detect the weak Stokes $V$ signal. To get a realistic estimate for a possible spectro-polarimetric observation of \ion{C}{4}, the estimate for the exposure time from \sect{S:required.exp} and \fig{F:intensity.noise} adapted for specifications similar to SUMER are used. In particular, a resolution element of about 2\arcsec\ and an exposure time of 12 minutes will be adopted. This can be considered as a worst case, because more modern instruments (will) have a significantly higher efficiency. If a Stokes $V$ signal is visible with such a long exposure, clearly it will be visible for the shorter exposure times of more modern instruments.

\begin{figure}
\centerline{\includegraphics{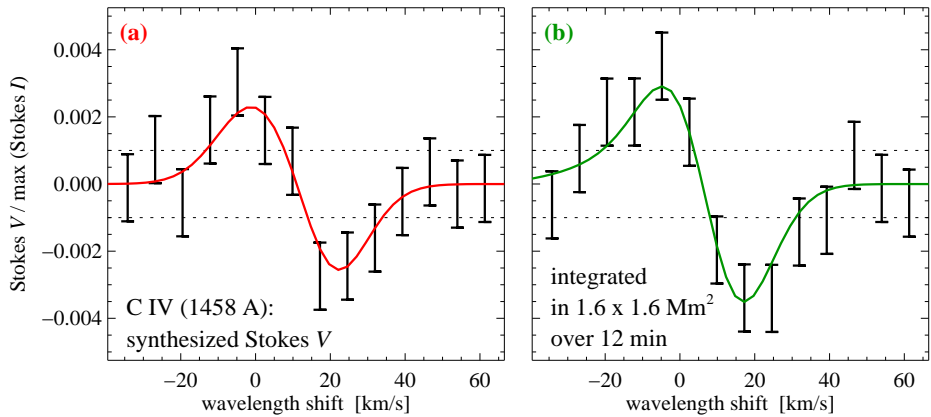}}
\caption{Realistic synthetic Stokes $V$ observation. The two panels show the synthesized Stokes $V$ profiles in the 1.6\,${\times}$\,1.6\,Mm$^2$ squares labeled in \fig{F:stokes.maps} accordingly. The synthetic observation was integrated in space over the square and in time for 12\,min. The bars show a potential observation at a spectral sampling comparable to that of the SUMER instrument with a noise level of $10^{-3}$.
See \sect{S:real.obs}.}
\label{F:stokes.profiles}
\end{figure}

Figure \ref{F:stokes.profiles} shows synthesized Stokes $V$ profiles integrated in space over a square of 1.6$\times$1.6\,Mm$^2$ (boxes in \fig{F:stokes.maps}) and in time over the 12 minutes for which the temporal variation of the intensity is shown in these squares in \fig{F:stokes.vari}. To get a more realistic representation, the Stokes $V$ spectra are shown as bars with a spectral sampling comparable to the SUMER instrument and with the addition of a noise level of 10$^{-3}$.

This shows that in the both regions selected here, close to the pore and in its vicinity, one can detect a Stokes $V$ signal with an instrument having a detection limit in Stokes $V$ over $I$ of about 10$^{-3}$. The main result from this experiment is that a Stokes $V$ signal survives even when integrating in space and time, despite the spatial and temporal fluctuations of the intensity on scales shorter than the length and time scales the spectrum is integrated during the exposure. This is basically because the magnetic field is comparatively constant (in time and space), at least much smoother than the intensity (see \sect{S:synt.vari}).
In conclusion, this model shows that the question posed in the heading of the current section can be answered with: yes, we can afford long exposure times to get information on the magnetic field.

It might be that one can carry over this conclusion to existing measurements of the magnetic field in the upper atmosphere by spectro-polarimetry in the infrared. For example the signals of forbidden and optically-thin lines detected in active region coronal loops with a coronagraph \citep[\textit{e.g.}][]{Lin+al:2000} might represent the background magnetic field. The polarization signals of the allowed He {\sc i} 10830\,\AA\ triplet, which is not optically thin, observed in spicules above the limb \citep{Centeno+al:2010} may well represent the magnetic field of the spicular plasma itself, as argued by these authors.

\section{How reliable is an inversion of the transition region magnetic field?\label{S:inversion}}

Even if one can detect a Stokes $V$ signal in \ion{C}{4}, it is not clear to what extent one can invert the longitudinal component of the magnetic field in the source region of the transition region emission. To test this, as a first step we derive a very simple magnetograph equation from \eqn{E:stokes} and compare the resulting inverted magnetic field to the magnetic field in the 3D model. Using the wavelength and the effective Land\'e factor for the \ion{C}{4} line, $\lambda{=}1548$\,\AA\ and $\bar{g}{=}1.167$, and assuming that the line width is comparable to the thermal width, $\Delta\lambda{\approx}0.06$\,{\AA}, one can rewrite \eqn{E:stokes} as
\begin{equation}\label{E:magnetograph}
B_{\rm{me}}~[G] ~\approx~ 46\,000~\overline{V} ~.
\end{equation}
Because in the 3D model the magnetic field $B$ is known at each grid point, one can now compare the inverted magnetic field $B_{\rm{me}}$ to the actual magnetic field in the source region to test this simplest inversion procedure.

As a first step we investigate the two realistic sample spectra shown in \fig{F:stokes.profiles}. Here the Stokes $V$ signals for regions (a) and (b) are $\overline{V}\approx{0.45}$\% and $0.55$\%. Following \eqn{E:magnetograph} this corresponds to inverted magnetic field strengths of $B_{\rm{me}}\approx{200}$\,G and 250\,G. These inverted values are plotted as dashed lines in \fig{F:stokes.vari} on top of the variation of the magnetic field during the exposure of these spectra. In these two cases the simple inversion seems to work and represents some average value of the (line-of-sight) magnetic field in the source region of the transition region emission. Comparing the actual and inverted magnetic field in the bottom panels of \fig{F:stokes.vari} we estimate the error of this procedure to be about 20\%.

\begin{figure}
\centerline{\includegraphics{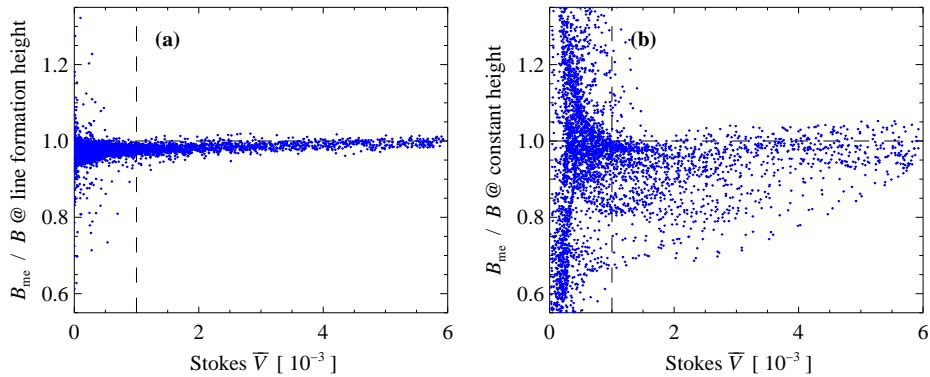}}
\caption{Relation of a simple magnetic field reconstruction to the synthesized Stokes signal $\overline{V}$ of \ion{C}{4} as defined in \eqn{E:stokes.map}. Panel (a) shows the ratio of the magnetic field strength $B_{\rm{me}}$ reconstructed by the magnetograph equation \eqn{E:magnetograph} to the magnetic field at the line formation height, \textit{i.e.}, on the currogated source surface of \ion{C}{4}.
In contrast, panel (b) shows the ratio of $B_{\rm{me}}$ to the magnetic field at a fixed height ($z\,{=}\,3.3$\,Mm), where the horizontally averaged temperature is $10^5$\,K.
See \sect{S:inversion}.}
\label{F:inversion}
\end{figure}

To test if the simple inversion also works in a statistical sense, we look at the scatter of the ratio of the inverted field $B_{\rm{me}}$ to the actual (line-of-sight) magnetic field $B$ in the \ion{C}{4} source region. This is shown in \fig{F:inversion}a as a function of the Stokes $V$ signal. At least in those areas where the Stokes $V$ signal is above a noise level of $10^{-3}$, there is a close correlation between the inverted and the actual magnetic field: the inverted signal is within about 5\% of the value from the 3D model. However, it has to be remembered, that this is without any noise, which will (in part) degrade the nice correlation.

From \fig{F:stokes.height}a it is clear that the \ion{C}{4} source region is far from being close to a flat plane. Of course, this raises the question on the usability of the inverted values of the magnetic field, which mainly reflects the field on the highly corrugated source surface of the transition region emission in \ion{C}{4} as seen in \fig{F:inversion}a. Therefore in \fig{F:inversion}b we compare the inverted magnetic field $B_{\rm{me}}$ to the magnetic field at a constant height in the computational box. Here we choose a height of $z{=}3.3$\,Mm, which in this model represents the height where the \emph{horizontally averaged} temperature reaches $10^5$\,K. This shows a much stronger scatter. Still, if one assumes that the measurement would represent the magnetic field at a constant height, one would be able to invert the magnetic field within some 20\% to 30\% (most values in \fig{F:inversion}b above a Stokes $V$ signal of 0.1\% have a ratio from 0.75 to 1.05). Of course, this is again without assuming noise of the measurement.

Certainly, the community will have to develop more elaborate inversion procedures to interpret the Stokes $V$ measurements in \ion{C}{4} in the future. However, even with the simple magnetograph-type inversion presented here one can hope to measure the transition region magnetic field within some 30\%.

\section{Conclusions\label{S:conclusions}}

This paper presents a forward modeling of the Stokes profiles in the extreme UV to investigate the diagnostic potential of emission lines for investigations of the magnetic field in the transition region and low corona. This study shows that with instruments providing a polarimetric sensitivity of Stokes $V/I$ of about 0.1\% this task to measure the transition region magnetic field can be reached; probably within some 20\% to 30\% using simple magnetograph-type inversions as a first step.

This study for \ion{C}{4} employs a realistic 3D MHD model that self-consistently solves for the magnetic structure \emph{and} the transition region and coronal plasma properties. The emission  synthesized from the model is highly structured in space and shows a strong temporal variability, just as actual observations. Naively one would expect that this high level of variation would destroy any significant polarization signal from the magnetic field. However, because the magnetic field structure is rather stable and smooth, the Stokes $V$ signal survives even when integrating in space and time.

Therefore spectro-polarimeters operating in the extreme UV will be able to provide reliable diagnostics of the transition region magnetic field. During its first flight in summer 2010 the SUMI rocket acquired some minutes worth of polarization data in \ion{Mg}{2} \citep{West+al:2011.polarization.full}. A second flight took place in summer 2012. The \ion{C}{4} data of this latter flight are currently calibrated and analysed, and it will be of high interest to see how they compare to the synthesized model data shown in this study --- in particular if the field-of-view of SUMI would include some regions comparable to the magnetic structures shown here.

Of course, it would be desirable to have measurements of the magnetic field in the transition region as proposed here on a space-based observatory, and embedded into a suite of instruments that will provide information on the magnetic field also in the chromosphere and the corona. The SolmeX space mission \citep{Peter+al:2012.solmex.full} would have provided such comprehensive measurements of the coronal magnetic field in the solar upper atmosphere, and was proposed to ESA recently. Continuing the theoretical and instrumental efforts, an opportunity might open up to confront the synthetic Stokes $V$ data from the transition region presented here to actual observations.

%
\begin{acks}
The author greatly acknowledges the collaboration with B.\,Gudiksen and \AA.\,Nordlund that stood at the beginning of this project to derive synthetic observations from coronal models. In particular, the author is grateful to B.\,Gudiksen for sharing data used in this study. Sincere thanks are due to J. Trujillo Bueno for his comments on the manuscript.
\end{acks}

%
\def\aap    {Astron. Astrophys.}
\def\aaps   {Astron. Astrophys. Suppl.}
\def\apj    {Astrophys. J.}
\def\apjl   {Astrophys. J. Lett.}
\def\apjs   {Astrophys. J. Suppl.}
\def\solphys{Solar Phys.}
\bibliographystyle{spr-mp-sola}


\end{article} 
\end{document}